In situ etching of *β*-Ga$_2$O$_3$ using *tert*-butyl chloride in an MOCVD system

Cameron A. Gorsak,[1] Henry J. Bowman,[2] Katie R. Gann,[1] Joshua T. Buontempo,[1] Kathleen T. Smith,[3] Pushpanshu Tripathi,[1] Jacob Steele,[1] Debdeep Jena,[1,4,5] Darrell G. Schlom,[1,5,6] Huili Grace Xing,[1,4,5] Michael O. Thompson,[1] and Hari P. Nair[1a)]

[1] Department of Materials Science and Engineering, Cornell University, Ithaca, New York 14853, USA
[2] PARADIM REU, Cornell University, Ithaca, New York 14853, USA
[3] School of Applied and Engineering Physics, Cornell University, Ithaca, New York 14853, USA
[4] School of Electrical and Computer Engineering, Cornell University, Ithaca, New York 14853, USA
[5] Kavli Institute at Cornell for Nanoscale Science, Cornell University, Ithaca, New York 14853, USA
[6] Leibniz-Institut für Kristallzüchtung, Max-Born-Str. 2, 12489 Berlin, Germany

[a)] Author to whom correspondence should be addressed: hn277@cornell.edu

In this study, we investigate *in situ* etching of *β*-Ga$_2$O$_3$ in a metal-organic chemical vapor deposition (MOCVD) system using *tert*-Butyl chloride (TBCl). We report etching of both heteroepitaxial ($\bar{2}$01)-oriented and homoepitaxial (010)-oriented *β*-Ga$_2$O$_3$ films over a wide range of substrate temperatures, TBCl molar flows, and reactor pressures. We infer that the likely etchant is HCl (g) formed by the pyrolysis of TBCl in the hydrodynamic boundary layer above the substrate. The temperature dependence of the etch rate reveals two distinct regimes characterized by markedly different apparent activation energies. The extracted apparent activation energies suggest that at temperatures below ~800 °C the etch rate is likely limited by desorption of etch products. The relative etch rates of heteroepitaxial ($\bar{2}$01) and homoepitaxial (010) *β*-Ga$_2$O$_3$ were observed to scale by the ratio of the surface energies indicating an anisotropic etch. Relatively smooth post-etch surface morphology was achieved by tuning the etching parameters for (010) homoepitaxial films.

The ultra-wide bandgap semiconductor *β*-Ga$_2$O$_3$ (~4.8 eV) has garnered attention recently as a platform for power electronics and radio frequency devices.[1] The ultra-wide bandgap results in a high critical breakdown field strength yielding a superior Baliga's figure of merit relative to semiconductors like SiC and GaN.[2] Progress in *β*-Ga$_2$O$_3$ research has been spurred by the availability of large-area (up to 4 inch) melt-grown substrates[3] and the ease of n-type doping.[4] Metal-organic chemical vapor deposition (MOCVD) has emerged as a technique capable of producing high-quality *β*-Ga$_2$O$_3$ thin films with room-temperature electron mobilities approaching the polar optical phonon limit.[5-8] A low-damage *in situ* etch to minimize contamination or plasma-induced damage before subsequent deposition of n+ material[9] or dielectrics[10] will be key for enabling higher-performance devices. In this study, we investigate using *tert*-Butyl chloride (TBCl) as a precursor for *in situ* etching of *β*-Ga$_2$O$_3$.

*In situ* etching of *β*-Ga$_2$O$_3$ has been demonstrated using a flux of elemental gallium in molecular beam epitaxy (MBE) and using triethylgallium in MOCVD.[11,12] The etch mechanism for both leverages the formation of volatile gallium suboxides.[13] The use of elemental Ga can, however, potentially leave gallium metal droplets on the surface necessitating an *ex situ* HCl wet etch. Agnitron Technologies has demonstrated that these Ga droplets can be removed *in situ* with TBCl etching, however, their use of TBCl for etching of $\beta$-Ga$_2$O$_3$ itself was not promising, requiring much higher TBCl molar flow compared to those used in this work to achieve appreciable etch rates.[14] *In situ* etching of $\beta$-Ga$_2$O$_3$ has also been demonstrated in halide vapor phase epitaxy (HVPE) systems using HCl gas.[15,16]

TBCl is an attractive choice as an etchant precursor since it is relatively noncorrosive compared to HCl, displays long-term stability at room temperature, has a reasonable vapor pressure, and can be installed in the typical bubblers, widely used for precursor delivery in MOCVD systems. *In situ* etching using TBCl has been demonstrated for various III-V semiconductors.[17-21] The thermal decomposition of TBCl has been experimentally confirmed to follow:[22-24]

$$(CH_3)_3CCl(g) \rightarrow iso\text{-}C_4H_8(g) + HCl(g). \qquad (1)$$

*In situ* etching of $\beta$-Ga$_2$O$_3$ was performed in a cold-wall Agnitron Agilis 100 MOCVD system equipped with a remote-injection showerhead. The SiC-coated graphite susceptor was inductively heated and the substrate temperature was measured by a pyrometer aimed at the backside. The etching studies were carried out over reactor pressures of 10 – 60 Torr, susceptor temperatures of 700 – 1000 °C, and TBCl molar flows of ~20 – 61 µmol/min. The stainless-steel bubbler containing the TBCl (99.9999%) was purchased from Dockweiler Chemicals and was held at a pressure of 900 Torr and a temperature of 5 °C. Both heteroepitaxial ($\bar{2}$01) $\beta$-Ga$_2$O$_3$, grown on c-plane sapphire, and homoepitaxial films, grown on Fe-doped (010) $\beta$-Ga$_2$O$_3$ from Novel Crystal Technologies (NCT), were etched. For homoepitaxial films, etching was studied at 15 and 30 Torr, 750 and 875 °C, and a TBCl molar flow of ~61 µmol/min. During etching, the total flow in the reactor was fixed at 6000 sccm using Ar (99.999%) carrier gas.

UV-vis optical reflectometry was used to measure heteroepitaxial film thickness, while X-ray diffraction (XRD) (PANalytical Empyrean) was used to determine homoepitaxial film thickness. Homoepitaxial films grown for determining the etch rate included a thin (~10 nm) $\beta$-(Al$_{0.07}$Ga$_{0.93}$)$_2$O$_3$ interface followed by 200-300 nm of $\beta$-Ga$_2$O$_3$ which provided an index contrast resulting in Laue oscillations.[25] Atomic force microscopy (AFM) was used to evaluate the surface morphology.

We used heteroepitaxial $\beta$-Ga$_2$O$_3$ to map out the etch rate as a function of TBCl molar flow at temperatures between 700 and 900 °C (Fig. 1). At a fixed reactor pressure of 15 Torr, we found that etch rate increases linearly with TBCl molar flows between ~20 – 61 µmol/min, which enables fine control of the etch rate.

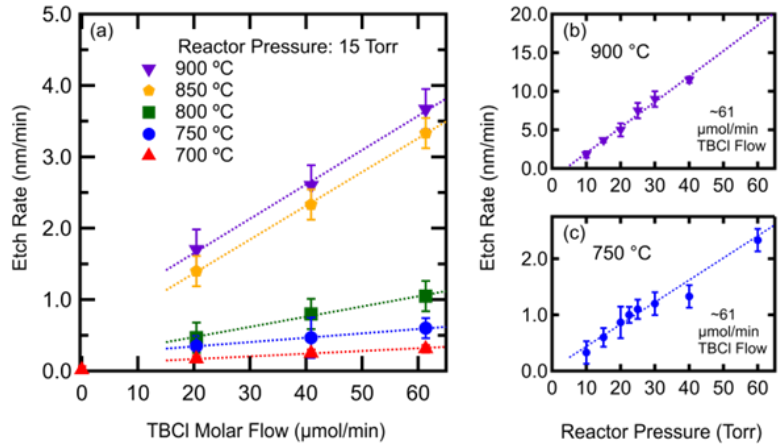

FIG. 1. Etch rate of heteroepitaxial $\beta$-Ga$_2$O$_3$ as a function of (a) TBCl molar flow and (b-c) reactor pressure. (a) At a fixed pressure of 15 Torr and temperatures between 700 and 900 °C, increasing the TBCl molar flow results in a linear increase in etch rate. At a fixed TBCl molar flow of ~61 µmol/min, the etch rate monotonically increases with reactor pressure in both the (b) HT and (c) LT regimes indicating that TBCl etching of $\beta$-Ga$_2$O$_3$ does not occur in a mass-transport-limited regime.

In general, the etch rate increases with increasing temperature, however, the slope of the etch rate vs TBCl molar flow jumps sharply from 800 °C to 850 °C, indicative of a sudden change in the etch-limiting step. At a fixed TBCl molar flow of ~61 µmol/min, the etch rate follows an Arrhenius relationship (Fig. 2). There are two distinct activation energy regimes which are commonly observed in CVD growth[26] and etching[27] processes. In our work, the low-temperature (LT) regime below ~800 °C has a higher apparent activation energy of ~1.59 and ~1.75 eV for 15 and 30 Torr, respectively. The high temperature (HT) regime above ~800 °C exhibits a much lower apparent activation energy of ~0.11 and ~0.04 eV for 15 and 30 Torr, respectively. These values are much lower than the Ga-O bond dissociation energy (~3.88 eV)[28] and we confirmed that even at the highest etch temperature employed there is negligible thermal decomposition of $\beta$-Ga$_2$O$_3$.[29]

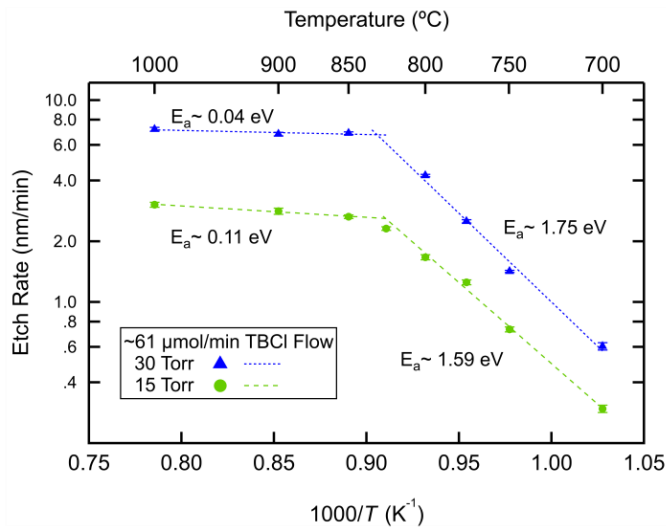

FIG. 2. Arrhenius plot of *in situ* etch rate of heteroepitaxial *β*-Ga$_2$O$_3$ for a TBCl molar flow of ~61 μmol/min at reactor pressures of 15 and 30 Torr. The HT and LT etch regimes are delineated by distinct apparent activation energies.

During the etching process, the etchant adsorbs on the surface and then reacts to form an etch product, followed by the desorption of the etch product from the surface. Based on the data from Tsang,[24] at temperatures between 700 -1000 °C used in our work and estimated residence times[30] for precursors in the heated boundary layer above the susceptor, TBCl pyrolyzes into isobutene and hydrogen chloride (HCl) (Fig. 3). This temperature range, however, is not high enough to enable further gas-phase pyrolysis of HCl (Fig 3).[31] Therefore, it is reasonable to assume HCl is the etchant.

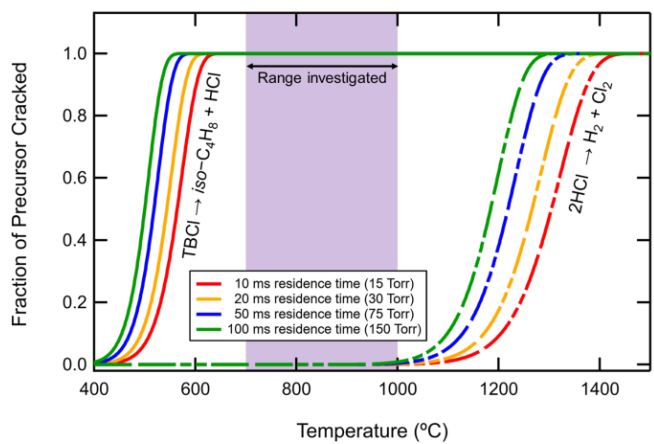

FIG. 3. Calculated fraction of TBCl and HCl pyrolyzed as a function of temperature based on typical residence times for precursors in the hot boundary layer above the substrate. The pyrolysis of TBCl into isobutene and hydrogen chloride (HCl) is fully complete by 600 °C, while the thermal decomposition of HCl does not occur until over 1000 °C.

We hypothesize that the HCl adsorbs onto the surface of the $\beta$-$Ga_2O_3$ and reacts to form volatile $GaCl_n$ ($n \leq 3$) etch product by one of the following three reactions which were determined to be thermodynamically favorable based on thermochemical data[32-34] and the partial pressures of the various species at experimental conditions used in this work:

$$Ga_2O_3(s) + 6HCl(g) \leftrightarrow 2GaCl(g) + 3H_2O(g) + 2Cl_2(g), \quad (2)$$
$$Ga_2O_3(s) + 6HCl(g) \leftrightarrow 2GaCl_2(g) + 3H_2O(g) + Cl_2(g), \quad (3)$$
$$Ga_2O_3(s) + 6HCl(g) \leftrightarrow 2GaCl_3(g) + 3H_2O(g). \quad (4)$$

In the LT regime, the weak dependence of etch rate on TBCl molar flow (Fig. 1) suggests the etch is limited by etch product desorption – either $GaCl_n$ or $H_2O$. While all thermodynamically favorable gas phase reactions can occur simultaneously, $GaCl_3$ formation kinetics will depend on the surface coverage of HCl, which is likely low given the linear (first order) increase in etch rate with TBCl partial pressure, and the high vapor pressure of HCl even at 700 °C.[35,36] Therefore, the LT etch product is likely GaCl. Surface science studies of $GaCl_n$ desorption from GaAs have shown an apparent activation energy for the desorption of $GaCl_3$ and GaCl of ~0.78 eV and ~1.65 eV,[37] respectively.[38] When the apparent activation energy is greater than that expected for the relevant $GaCl_n$ species, the limiting factor has been attributed to the anionic species (N for GaN and As for GaAs).[39,40] For the case of $\beta$-$Ga_2O_3$, after the desorption of volatile $GaCl_n$ species, the surface is likely left hydroxylated. The hydroxyl-terminated $\beta$-$Ga_2O_3$ surface is observed to be stable to 750 °C in an ultra-high vacuum (UHV) environment.[41] Although the activation energy for the desorption of $H_2O$ from hydroxylated $\beta$-$Ga_2O_3$ is unknown at this time, the apparent activation energy for the desorption of $H_2O$ from $\alpha$-$Al_2O_3$ surfaces is in the range of ~1.37 - 1.78 eV, which is similar to our observations in the LT regime.[42,43] Further studies are required to fully elucidate the LT etch mechanism.

In the HT regime, the dominant $GaCl_n$ species is presumably GaCl based on thermodynamic calculations of gas phase HVPE growth of GaN.[44] The low apparent activation energy in the HT regime agrees with the apparent activation energy (~0.08 eV) extracted above 800 °C from atmospheric pressure HCl etching of $\beta$-$Ga_2O_3$.[16] The etch rate in the HT regime is determined by the surface concentration of HCl as evidenced by the linear dependence of the etch rate on reactor pressure in Fig. 1(b).[45]

At high substrate temperatures, MOCVD growth typically occurs in a mass-transport-limited regime. In the absence of significant gas-phase parasitic reactions, and when all the flows are held constant, the growth rate is independent of total reactor pressure within this mass-transport-limited regime.[30,46] To determine whether etching with TBCl occurs in a mass-transport-regime, the pressure dependence of the etch rate was explored at a fixed TBCl flow rate at two temperatures: 750 °C (LT regime) and 900 °C (HT regime). The reactor pressure was controlled, independent of the total gas flow, using a computer-controlled butterfly valve in the exhaust manifold. Figures 1(b) and 1(c) show that the etch rate is increasing, and not saturating, with increasing reactor pressure indicating that etching is not occurring in a mass-transport-limited regime even at the highest substrate temperatures investigated in this study. An increase in etch rate with reactor pressure is likely due to the increased surface coverage of the etchant, HCl, with increasing HCl partial pressure.[45]

In order to investigate the effect of oxygen on etch rate we introduced a 50 sccm flow of $O_2$ into the reactor during etching. In the LT regime, the etch rate was largely not affected, however, in the HT regime the etch rate was suppressed by a factor of ~2 as seen in Fig. 4. We believe that the HT etch rate is suppressed due to the competing HVPE growth back reaction.

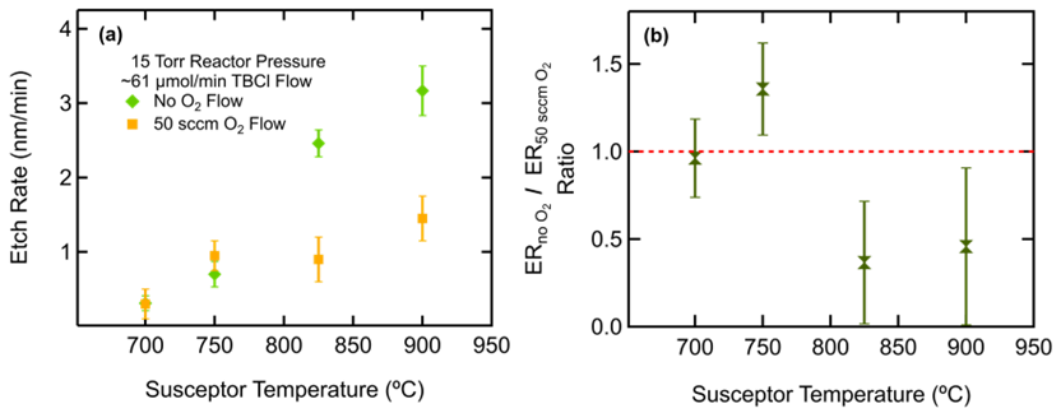

FIG. 4. Effect of additional $O_2$ flow on etch rate. (a) Etch rate as a function of susceptor temperature with no $O_2$ (green) and with 50 sccm $O_2$ flow (yellow). (b) Ratio of the etch rate without oxygen divided by the etch rate with oxygen vs susceptor temperature. Below 800 °C the etch rate is relatively unchanged but is suppressed at higher temperatures.

Next, we etched co-loaded heteroepitaxial ($\bar{2}01$) and homoepitaxial (010) $\beta$-Ga$_2$O$_3$ samples to investigate the etch rate anisotropy. The etch rate for homoepitaxial samples with ~61 μmol/min TBCl molar flow was determined by XRD (Fig. S1) for four distinct conditions (15 Torr at 750 and 875 °C, and 30 Torr at 750 and 875 °C). Figure 5 summarizes these etch rates revealing the anisotropy. Also plotted in Fig. 5 are the ratios between the calculated (010) and ($\bar{2}01$) dangling bond densities ($\rho$, green) and surface energies ($E$, blue) which agrees well with the experimental data.[47] Etch rate anisotropy has also been observed during etching of $\beta$-Ga$_2$O$_3$ using HCl in an HVPE reactor.[15]

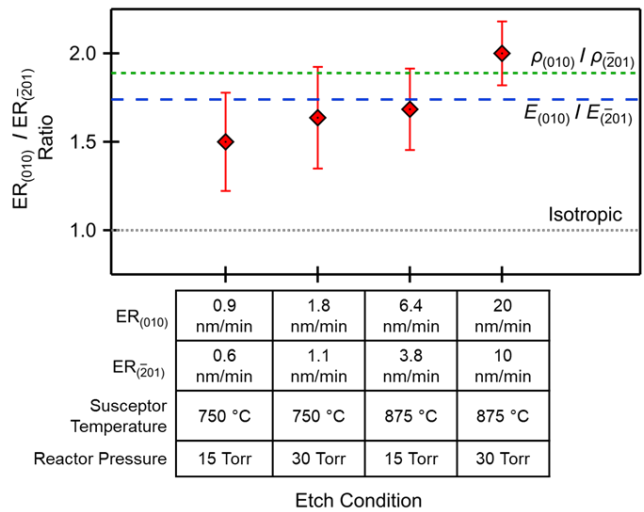

FIG. 5. Comparison of etch rate between homoepitaxial (010) and heteroepitaxial ($\bar{2}01$) $\beta$-Ga$_2$O$_3$ with ~61 μmol/min TBCl. Included in the plot are the ratios of calculated $\beta$-Ga$_2$O$_3$ dangling bond densities ($\rho$, green) and surface energies ($E$, blue) from Mu *et al.* (2020) illustrating that the anisotropy of the etch rate is correlated with the surface energy anisotropy: the higher the surface energy, the higher the etch rate.

To investigate the surface morphology resulting from TBCl etching, ~400 nm thick homoepitaxial (010) unintentionally doped (UID) $\beta$-Ga$_2$O$_3$ samples were grown and then immediately *in situ* etched, without cooling down, to a depth of ~100 nm for each of the four conditions shown in Fig. 6(b-e). The resulting surface morphology resembles that resulting from hot phosphoric wet etching[48] and does not exhibit characteristic faceting of the (110) plane along the [001] direction typically seen post-growth[49] (as shown for an unetched film in Fig. S1(b)) or after elemental Ga etching.[11,12] In general we observe that etching under higher pressures and low temperatures results in smoother surfaces. Currently, the exact mechanism for surface roughening is unclear but we note that conditions for smoother etch morphologies also result in longer surface residence time of gas-phase species. We note that heteroepitaxial ($\bar{2}01$) oriented β-Ga$_2$O$_3$ films grown on c-plane sapphire also exhibited increased surface roughness after etching using TBCl (Fig. S2).

The electrical properties are not compromised for films grown after a 30 minute *ex situ* 48% hydrofluoric acid etch plus a ~50 nm *in situ*, ~61 μmol TBCl etch at 750 °C. (010) homoepitaxial films doped[50] to ~1 x 10$^{17}$ and ~2 x 10$^{18}$ cm$^{-3}$ exhibit mobilities of ~115 and ~102 cm$^2$/Vs, respectively. We demonstrate that subsequent regrowth after *in situ* etching of the substrate results in sub-nanometer RMS roughness (Fig. S3). The surface morphology of regrowth after *in situ* etching homoepitaxial films is comparable (~1 nm RMS) and is the subject of future work.

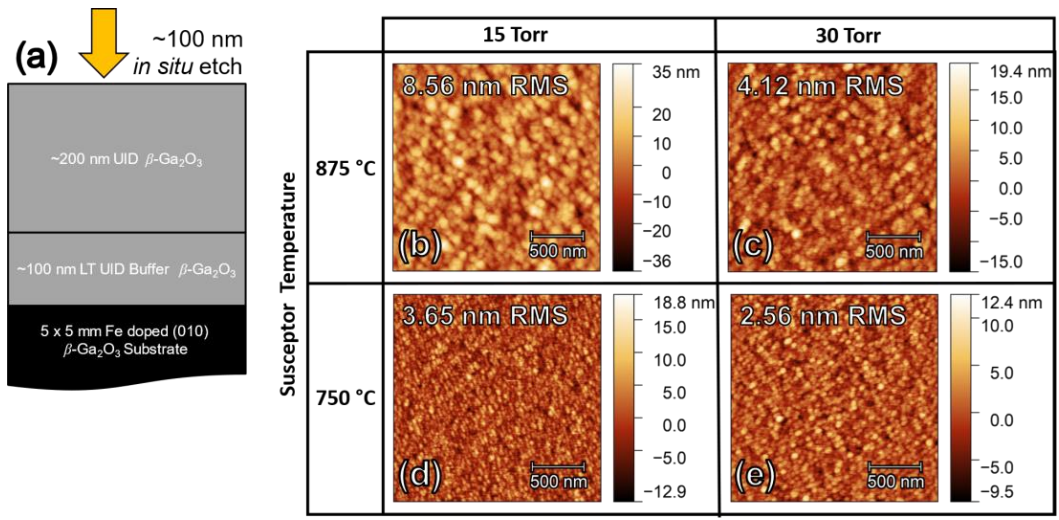

FIG. 6. Surface morphologies of *in situ* TBCl etched β-Ga$_2$O$_3$. (a) Layer structure illustrating ~100 nm *in situ* etch after growth of ~400 nm of UID β-Ga$_2$O$_3$ on Fe-doped (010) substrates. (b-e) AFM of surfaces after etching with ~61 μmol/min TBCl molar flow at various conditions. The 750 °C etch (bottom row) is smoother than at 875 °C (top row). 30 Torr etching (right column) is smoother than 15 Torr (left column).

In summary, this study investigated the *in situ* etching of both heteroepitaxial ($\bar{2}01$) and homoepitaxial (010) β-Ga$_2$O$_3$ films by TBCl in an MOCVD system over a temperature range of 700 – 1000 °C, pressure of 10 – 60 Torr, and TBCl molar flow of ~20 – 61 μmol/min. Two distinct regimes for TBCl etching of β-Ga$_2$O$_3$ were observed. The LT regime, below ~800 °C, exhibits an apparent activation energy of ~1.59 and ~1.75 eV for 15 and 30 Torr, respectively. In the LT regime, we hypothesize the etch rate is limited by the desorption of GaCl$_n$ or likely H$_2$O. In the HT regime, we hypothesize the thermodynamically favored etch product is GaCl and the apparent activation energy is low. The relationship between the etch rate of ($\bar{2}$01) and (010) β-Ga$_2$O$_3$ scales by the ratio of surface energies. Finally, the surface morphology of *in situ* etched homoepitaxial films was evaluated and it was determined that the lower temperature, higher pressure etch resulted in smoother surfaces. This work lays the groundwork for utilizing *in situ* TBCl etching and regrowth to obtain low-resistance ohmic contacts and improve the performance of β-Ga$_2$O$_3$ based devices.

Supplementary Material

The supplementary material includes the MOCVD growth conditions for the films etched in this study. Also shown is the XRD used for etch rate determination. AFM is included of as-grown and regrown homoepitaxial films as well as AFM of an as-grown and etched heteroepitaxial film.

Acknowledgments

We acknowledge support from the AFOSR/AFRL ACCESS Center of Excellence under Award No. FA9550-18-10529. C.A.G acknowledges support from the National Defense Science and Engineering Graduate (NDSEG) Fellowship. H.J.B. acknowledges support from the National Science Foundation (NSF) [Platform for the Accelerated Realization, Analysis and Discovery of Interface Materials (PARADIM)] under Cooperative Agreement No. DMR-1539918. We also acknowledge support from PARADIM for XRD usage. Substrate dicing and AFM were performed in the Cornell NanoScale Facility, a member of the National Nanotechnology Coordinated Infrastructure (NNCI), which is supported by the NSF (Grant No. NNCI-2025233).

Conflicts of Interest

The authors have no conflicts to disclose.

Author Contributions

C. A. Gorsak and H. J. Bowman contributed equally to this paper.

Data Availability

The data that support the findings of this study are available from the corresponding author upon reasonable request.

Supplementary Material

*In situ* etching of *β*-Ga$_2$O$_3$ using *tert*-butyl chloride in an MOCVD system

Cameron A. Gorsak,[1] Henry J. Bowman,[2] Katie R. Gann,[1] Joshua T. Buontempo,[1] Kathleen T. Smith,[3] Pushpanshu Tripathi,[1] Jacob Steele,[1] Debdeep Jena,[1,4,5] Darrell G. Schlom,[1,5,6] Huili Grace Xing,[1,4,5] Michael O. Thompson,[1] and Hari P. Nair[1]

[1]Department of Materials Science and Engineering, Cornell University, Ithaca, New York 14853, USA
[2]PARADIM REU, Cornell University, Ithaca, New York 14853, USA
[3]School of Applied and Engineering Physics, Cornell University, Ithaca, New York 14853, USA
[4]School of Electrical and Computer Engineering, Cornell University, Ithaca, New York 14853, USA
[5]Kavli Institute at Cornell for Nanoscale Science, Cornell University, Ithaca, New York 14853, USA
[6]Leibniz-Institut für Kristallzüchtung, Max-Born-Str. 2, 12489 Berlin, Germany

Homoepitaxial Growth Conditions and Etch Rate Determination

All layers were grown at a reactor pressure of 15 Torr. Triethylgallium (TEGa), triethylaluminum (TEAl), and O$_2$ were used as precursors. The LT UID buffers were grown at 600 °C with a typical TEGa and O$_2$ flow of ~38.5 µmol/min and 500 sccm, respectively. The 800 °C UID layers were grown with a typical TEGa and O$_2$ flow of ~77 µmol/min and 400 sccm, respectively. The growth of the heteroepitaxial samples was similar.

A thin (~10 nm) layer of *β*-(Al$_{0.07}$Ga$_{0.93}$)$_2$O$_3$ serves as an interface to separate the *β*-Ga$_2$O$_3$ film from the underlying substrate so that thickness fringes from the ~200-300 nm thick film of *β*-Ga$_2$O$_3$ can be seen in *θ*-2*θ* XRD scans (Fig. S1(c-f)). Laue oscillations were used to determine the thickness. The thin *β*-(Al$_{0.07}$Ga$_{0.93}$)$_2$O$_3$ layers were grown at 705 °C with a TEGa and O$_2$ flow of ~80 µmol/min and 200 sccm, respectively. The TEGa:TEAl input ratio was ~16.

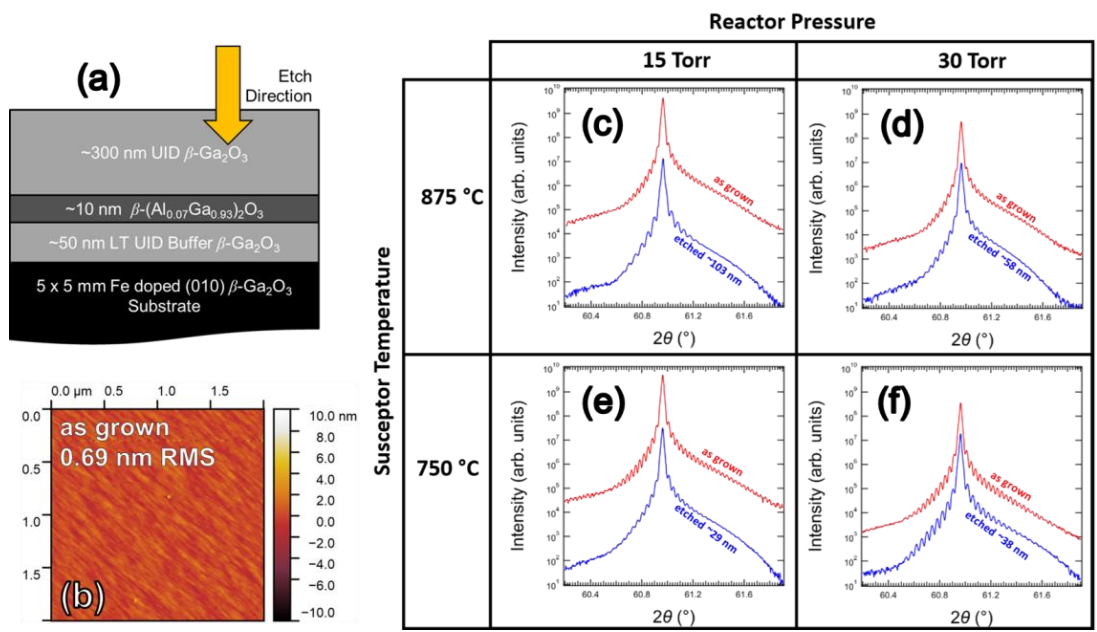

FIG. S1. Etch rate determination of homoepitaxial (010) *β*-Ga$_2$O$_3$. (a) Layer structure of as-grown sample. (b) AFM of an as-grown and not etched (010) *β*-Ga$_2$O$_3$ film surface indicating sub-nanometer RMS roughness (c-f) XRD of pre- and post-etched samples. Etching was done with ~61 µmol/min input TBCl molar flow at various conditions: 750 °C (bottom row), 875 °C (top row), 30 Torr (right column), and 15 Torr (left column). Laue oscillations were used to determine the thicknesses.

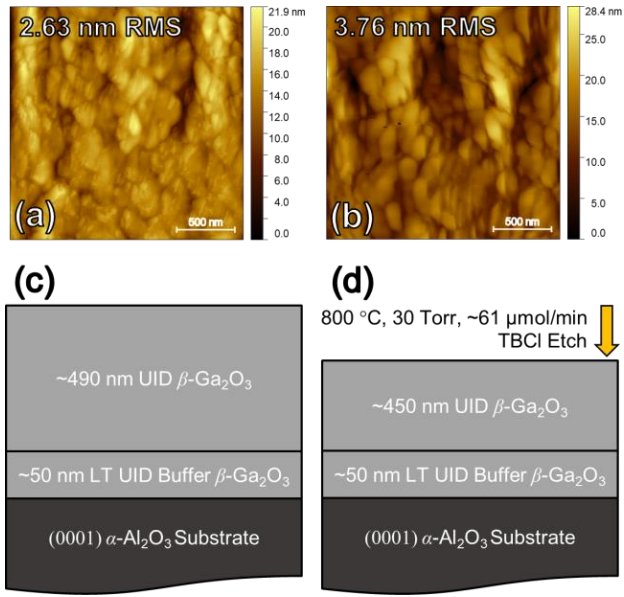

FIG. S2. Surface morphology of as-grown (a) and TBCl etched (b) heteroepitaxial ($\bar{2}01$) $\beta$-Ga$_2$O$_3$ with respective layer structure (c,d).

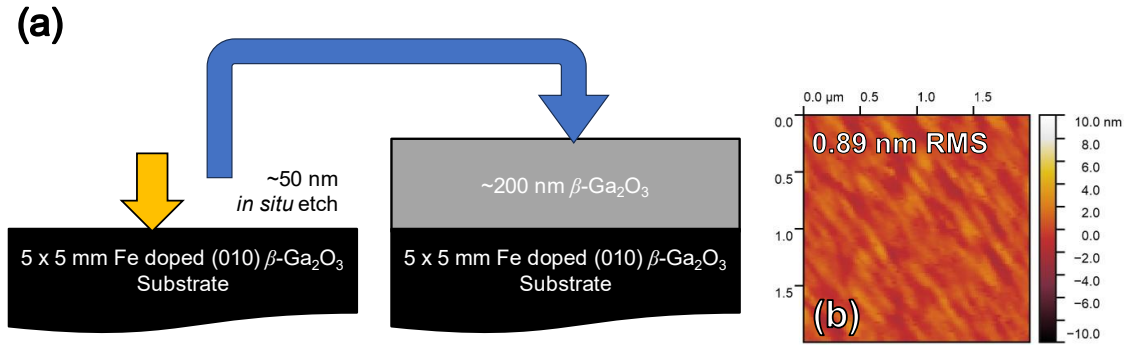

FIG. S3. (a) Layer structure of immediate regrowth after ~50 nm *in situ* TBCl etch with ~61 μmol/min input molar flow at 875 °C and 30 Torr. (b) AFM of the surface after regrowth showcasing sub-nanometer RMS roughness.